\documentclass[10pt,a4paper]{article}

\usepackage[utf8]{inputenc}
\usepackage[T1]{fontenc}
\usepackage[margin=2.0cm]{geometry}
\usepackage[numbered]{bookmark}
\usepackage{mathtools}
\usepackage{ulem}
\usepackage{amssymb}
\usepackage{graphicx}
\usepackage[font=small,labelfont=bf]{caption}
\usepackage{authblk}
\usepackage{cite}
\usepackage{dsfont}
\usepackage{xcolor}

\usepackage[titles]{tocloft}

\numberwithin{equation}{section}

\newcommand{\be}{\begin{equation}}
	\newcommand{\ee}{\end{equation}}
\newcommand{\bea}{\begin{eqnarray}}
	\newcommand{\eea}{\end{eqnarray}}

\newcommand{\nocontentsline}[3]{}
\newcommand{\tocless}[2]{\bgroup\let\addcontentsline=\nocontentsline#1{#2}\egroup}

\title{A review on the equation of state of the color superconductivity phase via holography in the Einstein-Gauss-Bonnet gravity}
\author[a,b]{Nguyen Hoang Vu
\footnote{Emails: \texttt{vu@jinr.ru} }}

\affil[a]{\textit{BLTP, JINR,}\textit{141980 Dubna, Moscow region, Russia}} 
\affil[b]{\textit{Institute of Physics, VAST,} \textit{10000, Hanoi, Vietnam}}

\date{}

\begin{document}
	\maketitle

	\begin{abstract}
		In this project, we will study a bottom-up holographic model  for the color superconductivity (CSC) phase in the Einstein-Gauss-Bonnet (EGB) gravity. We study the color superconductivity phase in the deconfinement phase which is dual to the planar GB-RN-AdS in six-dimensional spacetime and the confinement phase dual to the AdS soliton in Einstein-Gauss-Bonnet gravity and we find the equation of state of the CSC phase with both confinement and deconfinement phase.
	\end{abstract}


\section{Introduction}

The color superconductivity (CSC) phase is one of exotic phases in non-perturbative quantum chromodynaics (QCD). This is the condensate of two quarks in one Cooper pair in the low temperature (below the QCD scale) and high chemical potential (density), called the diquark, analogous to the Cooper pair in the metallic superconductivity.  Nevertheless, the strong interaction between two quarks is the attractive force and the diquark can be created directly, it is difference to the superconductivity in the metalic that the electron Cooper pair is paired by the interaction between electrons and phonon and the condensate of electron break the $U(1)_{em}$ only. In color superconductivity phase, the quark pairs carry the net color charge, the baryon number and the electric charge hence the condensation of the quark pairs breaks the $SU(3)_C$ gauge symmetry, $U(1)_B$ baryon symmetry and $U(1)_{em}$ symmetry spontaneously. And the Meissner effect of color supercondcutivity phase in QCD consits of the electromagnetic Meissner effect and the color Meissner effect. Moreover, the diquark can be created in the confinement phase, color superconductivity may be occur in the confinement matter \cite{Kazuo 2023}. And because the color superconductivity only occur when the chemical potential is large, we can claim that this phase is in the inner core of the massive neutron star or quark star \cite{Kazuo 2021},\cite{Kazuo 2023} and we can probe by observe the gravitational waves \cite{LIGO2017}.

One way to study the CSC phase is to apply the AdS/CFT correspondence or holography \cite{Maldacena98}. In AdS/CFT correspondence, one problem for strong interaction in quantum field theory (QFT) in $d$-dimension is dual to one problem in gravity in the $AdS_{d+1}$ with weak coupling constant. Hence, we can transform the problems of the color superconductivity phase to one problems of gravity theory with weak coupling constant in the AdS spacetime. The first papers who study color superconductivity via hholography is \cite{Basu 2011}. In \cite{Basu 2011}, the authors use the Einstein-Maxwell gravity and study the CSC phase in both confinement and deconfinement phase by the dual with AdS soliton and RN-AdS blackhole. And the diquark (in this paper, we only consider the scalar case of the diquark) corresponds to the scalar hair $\psi(r)$. Because the CSC phase occurs at low temperature (below the QCD scale), we add one compact extra dimension $y$ to the boundary that corresponds to the QCD scale and the scale of this compact extra dimension as $R_y$. Hence, the boundary this model becomes $R^{1,3}\times S^1$ and the bulk will be $AdS_6$. It is difference for holographic model for the metallic superconductivity \cite{Horowitz2008} when we use the $AdS_4$ to describe.

However, one problem is in \cite{Kazuo 2019} if we only use the Einstein-Maxwell gravity and standard Maxwell interaction, we only study the CSC with $N_c=1$ in the deconfinement phase and no CSC phase transition in confinement phase. To solve this problem, the author in \cite{nam2021more} proved that if we use Einstein-Gauss-Bonnet gravity, we can study CSC with $N_c=2$ and $N_c=3$ when $\alpha<0$ and the magnitude of $\alpha$ is sufficiently large with both confinement (very low temperature) and  deconfinement phase (finite temperature). In this paper, we will use this holographic model to probe the equation of state $p=p(\mu)$ of the color superconductivity phase in confinement and deconfinement case.

 The organization of this paper is as follows. In section \ref{sec2} and \ref{sec3} we quick review the holographic model for the CSC phase in EGB gravity and the ability for the CSC with $N_c=3$ in this model. In section \ref{sec4}, we find the form of equation of state of the CSC phase in confinement and deconfinement phase. And finally, in section \ref{sec5}, we conclude the main results and mention some open questions and interesting future directions.

\section{Holographic model for the color superconductivity in EGB gravity \label{sec2}}
In this section, first of all, we will quick review the $6d$ Einstein-Gauss-Bonnet gravity \cite{nam2021more}. The action of this dual model is:
\begin{equation}\label{AdS6 action}
S=\frac{1}{2k^2_6}\int d^6x\sqrt{-g}[R-2\Lambda+\tilde{\alpha}(R^2-4R_{\mu\nu}R^{\mu\nu}+R_{\mu\nu\rho\lambda}R^{\mu\nu\rho\lambda})+\mathcal{L}_{mat}],
\end{equation}
with the matter Lagrangian given by:
\begin{equation}
\mathcal{L}_{mat}=-\frac{1}{4}F_{\mu\nu}F^{\mu\nu}-|(\nabla_{\mu}-iqA_{\mu})\psi|^2-m^2|\psi|^2,
\end{equation}
where $\Lambda$ is the cosmological constant of the asymptotic AdS spacetime and this is related to the AdS radius $l$ as $\Lambda=-\frac{10}{l^2}$, the $\tilde{\alpha}=\frac{\alpha}{6}$ is the Gauss-Bonnet coupling parameter. In the matter part of this Lagrangian, the complex scalar field $\psi$ is dual to the diquark Cooper pair operator, the $U(1)$ gauge field $A_{\mu}$ corresponds to the current of the baryon number, and the $U(1)$ charge $q$ is regarded as the baryon number of the diquark. The baryon number of the diquark operator is related to the number of colors $N_c$ as $q=\frac{2}{N_c}$. And we set $1/2k_6^2=1$ and $l=1$ \cite{nam2021more}. And the ansatz for the vector and the scalar field: $A_{\mu}dx^{\mu}=\phi(r)dt$, $\psi=\psi(r)$  

In this model, the spacetime geometry dual to the deconfinement phase is the planar black hole. In Einstein-Gauss-Bonnet gravity the black hole solution:
\begin{equation}
ds^2=r^2(-f(r)dt^2+h_{ij}dx^i dx^j+dy^2)+\frac{dr^2}{r^2f(r)},
\end{equation}
where $h_{ij}dx^idx^j=dx_1^2+dx_2^2+dx_3^2$ is the line element of the $3$-d planar hypersurface and $y$ is the extra dimension that corresponds to the QCD scale and is compacted with the radius $R_y$. The blackening function of this configuration is as follows
\begin{equation}
f(r)=\frac{1}{2\alpha}\left[1-\sqrt{1-4\alpha\left(1-\frac{r^5_+}{r^5}\right)+\frac{3\alpha\mu^2}{2r^2_+}\left(\frac{r_+}{r}\right)^5\left(1-\frac{r^3_+}{r^3}\right)}\right].
\end{equation}
Where the bulk chemical potential $\mu$ is dual to the baryon chemical potential $|mu_B$ in the boundary

The temperatutre of this system in deconfinement phase corresponds to the Hawking temperature of the planar GB-RN-AdS black hole 
\begin{equation}
T=\frac{1}{4\pi}\left(5r_+-\frac{9\mu^2}{8r_+}\right).
\end{equation}

Using the nonnegative condition of the temperature, we have the constraint
\begin{equation}
    0\leq\frac{\mu}{r_+}\leq\frac{\sqrt{40}}{3}.
\end{equation}
From \cite{nam2021more} the $U(1)$ gauge field $\phi$ and the complex scalar field $\psi$ obeys:
\begin{equation}\label{eom}
\begin{split}
\phi''(r)+\frac{4}{r}\phi'(r)-\frac{2q^2\psi^2(r)}{r^2f(r)}\phi(r)&=0,\\
\psi''(r)+\left[\frac{f'(r)}{f(r)}+\frac{6}{r}\right]\psi'(r)+\frac{1}{r^2f(r)}\left[\frac{q^2\phi^2(r)}{r^2f(r)}-m^2\right]\psi(r)&=0.
\end{split}
\end{equation}

The matter fields (when $r\rightarrow\infty$) (because the boundary is $5d$ spacetime) are \cite{nam2021more}:
\begin{equation} 
\begin{split}
\phi(r)&= \mu-\frac{\overline{d}}{r^3}\\
\psi(r)&= \frac{C}{r^{\Delta_+}}+\frac{J_C}{r^{\Delta_-}}, 
\end{split}
\end{equation}
where $\mu$, $\overline{d}$, $J_C$, and $C$ are the chemical potential, charge density, source, and the condensate value (the VEV) of the bulk scalar field that is dual to the diquark Cooper pair,  respectively. The conformal dimension is given by
\begin{equation}
    \Delta_{\pm}=\frac{1}{2}(5\pm\sqrt{25+4m^2l^2_{eff}}),
\end{equation}
with $l^2_{eff}=\frac{2\alpha}{1-\sqrt{1-4\alpha}}$ and the Breitenlohner-Freedman (BF) bound \cite{BF1}, \cite{BF2} is
\begin{equation}
    m^2l^2_{eff}\geq -\frac{25}{4l^2_{eff}}.
\end{equation}

We assume that $m^2\ell^2_{eff}=-4$ we have at the boundary ($r\rightarrow\infty$)
\begin{equation}\label{boundary psi}
    \psi(r)= \frac{C}{r^4}+\frac{J_C}{r},
\end{equation}
and at the event horizon we have
\begin{equation}
\begin{split}
\phi(r_+)=0 \\
\psi(r_+)=r^2_+\frac{f'(r_+)\psi'(r_+)}{m^2}.
\end{split}
\end{equation}

And if we want to find the CSC phase transition in confinement phase, the spacetime gravity dual to the confinement phase is the GB-AdS soliton solution \cite{Cai 2002}, which is obtained via the analytic continuation from the planar GB-AdS black hole solution \cite{Cai 2007} as
\begin{equation}
\label{AdS soliton background}
ds^2=r^2\left(-dt^2+h_{ij}dx^idx^j+f(r)dy^2)+\frac{dr^2}{r^2f(r)}\right),
\end{equation}
where
\begin{equation}
f(r)=\frac{1}{2\alpha}\left[1-\sqrt{1-4\alpha\left(1-\frac{r_0^5}{r^5}\right)}\right],
\end{equation}
and $r_0=\frac{2}{5R_y}$, and the equation of motion with this configuration of spacetime as:
\begin{equation}
\label{EOM soliton}
\begin{split}
\phi''(r)+\left[\frac{f'(r)}{f(r)}+\frac{4}{r}\right]\phi'(r)-\frac{2q^2\psi^2(r)}{r^2f(r)}\phi(r)&=0\\
\psi''(r)+\left[\frac{f'(r)}{f(r)}+\frac{6}{r}\right]\psi'(r)+\frac{1}{r^2f(r)}\left[\frac{q^2\phi^2(r)}{r^2}-m^2\right]\psi(r)&=0
\end{split}
\end{equation}
And the boundary condition at $r_0$ is:
\begin{equation}
\begin{split}
\phi'(r_0)&=\frac{2q^2\psi^2(r_0)}{5r_0}\phi(r_0)\\
\psi'(r_0)&=-\frac{1}{5r_0}\left(\frac{q^2\phi^2(r_0)}{r_0^2}-m^2\right)\psi(r_0)
\end{split}
\end{equation}

\section{CSC phase with $N_c=3$ via holography  \label{sec3}} 
In this section, we review the ability to have the color superconductivity with $N_c=3$ in this holographic model in \cite{nam2021more}. At the critical chemical potential $\mu_c$, when the diquark is not formed yet, the back reaction of the scalar field $\psi$ on the dual background geometry is negligible. Thus the action  \eqref{AdS6 action} becomes
\begin{equation}\label{no matter action}
S_{no matter}=\int d^6x\sqrt{-g}\left[R-2\Lambda+\frac{\alpha}{6}(R^2-4R_{\mu\nu}R^{\mu\nu}+R_{\mu\nu\rho\lambda}R^{\mu\nu\rho\lambda})-\frac{1}{4}F_{\mu\nu}F^{\mu\nu}\right]
\end{equation}
The spacetime metric solution dual to the confinement phase is given by the AdS soliton solution in EGB gravity and the corresponding potential of the  gauge field as
\begin{equation}\label{gauge soliton solution}
\phi(r)=\mu
\end{equation}
Whereas, the spacetime metric solution dual to the deconfinement phase is given by the planar GB-RN-AdS black hole and the corresponding gauge field as:
\begin{equation}\label{gauge black hole solution}
\phi(r)=\mu\left(1-\frac{r_+^3}{r^3}\right)
\end{equation} 
In \cite{nam2021more}, we have the free energy of the planar GB-RN-AdS black hole and GB-AdS soliton as
\begin{equation}
\begin{split}
\Omega_{BH}&=-r_+^5\left(1+\frac{3\mu^2}{8r_+^2}\right)\frac{4\pi}{5r_0}V_3\\
\Omega_{Sol}&=-r_0^5\frac{4\pi}{5r_0}V_3
\end{split}
\end{equation}
The confinement-deconfinement phase diagram of this model is illustrated via Fig.1. The critical curve (red one) which separate the AdS soliton and AdS black hole in this model is determined by the equation $\Omega_{BH}=\Omega_{Sol}$
\begin{figure}[h!]
\centering
 \includegraphics[width=8.6cm]{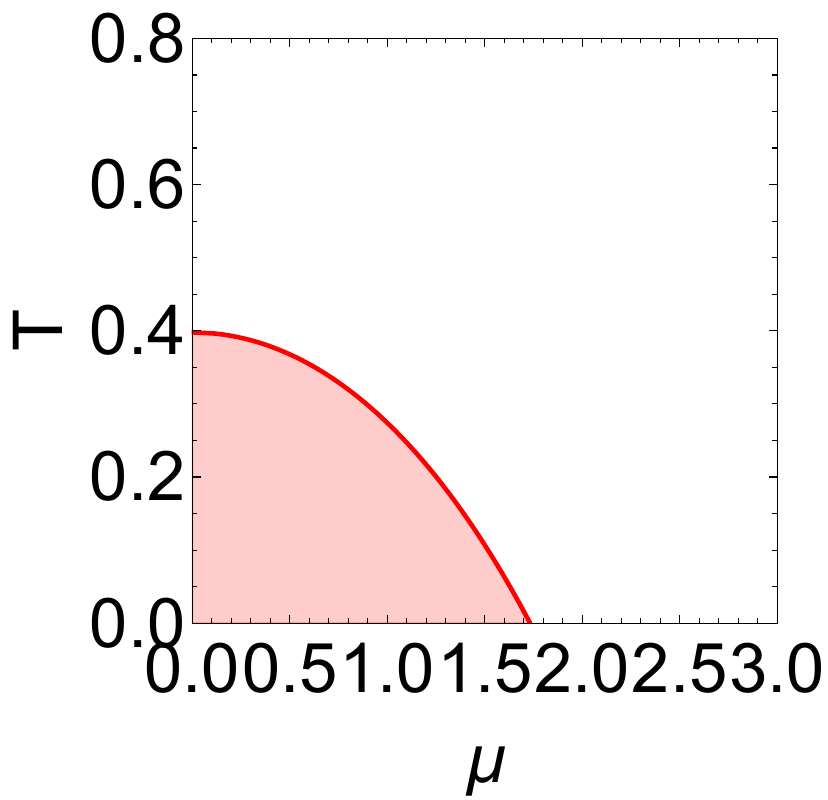}
 \label{fig:scalefactor}
 \caption{The confienement-deconfinement phase diagram for this holographic model. At $\mu=1.73$ we have the phase transition with T=0}
 \label{phase diagram}
\end{figure}
\subsection{CSC phase in deconfinement phase}
After introduce our holographic model, we discuss the ability to occur the color superconductivity in deconfinement phase. From the equation of motion in deconfinement, we have the effective mass 
\begin{equation}
m^2_{eff}=m^2-\frac{q^2\phi^2(r)}{r^2f(r)}
\end{equation}
This mass square describe the one difference scalar field in the $AdS_6$. To obtain the condensate of the diquarks in the boundary, we need the condensate of $\psi(r)$ in the bulk. And hence, we need instability of the field that associate with mass effective mass $m^2_{eff}$. Thus, the BF bound must to break with the effective mass, and we have
\begin{equation}
m^2_{eff}<-\frac{25}{4l^2_{eff}}
\end{equation}
and we have
\begin{equation}\label{instability condition}
\frac{q^2\phi^2(r)}{r^2f(r)}>\frac{9}{4l^2_{eff}}
\end{equation}
The left hand side of \eqref{instability condition} can be rewritten as
\begin{equation}
\frac{q^2\phi^2(r)}{r^2f(r)}=q^2\frac{2\alpha z^2(1-z^3)^2\hat{\mu}^2}{1-\sqrt{1-4\alpha(1-z^5)+\frac{3\alpha\hat{\mu}^2}{2}z^5(1-z^3)}}\equiv q^2\mathcal{F}(z,\hat{\mu},\alpha),
\end{equation}
where $z\equiv r_+/r$ ($0<z\leq 1$) and $\hat{\mu}\equiv\mu/r_+$ ($1.73\leq\hat{\mu}\leq\sqrt{40}/3$), because at $\hat{\mu}=1.73$ we have the confinement-deconfinement phase transition. Use the relation $q=2/N_c$ and we have
\begin{equation}
\frac{9}{4l^2_{eff}}<\frac{4}{N_c^2}\mathcal{F}(z,\hat{\mu},\alpha)
\end{equation}
After some manipulations, we obtain the condition of $N_c$ in the deconfinement phase
\begin{equation}
N_c<\frac{4}{3}\sqrt{\frac{2\alpha\mathcal{F}_{max}(z,\hat{\mu},\alpha)}{1-\sqrt{1-4\alpha}}}
\end{equation}
where $0<z\leq 1$ and $1.73\leq\hat{\mu}\leq\sqrt{40}/3$. In \cite{nam2021more}, the author proved that with $\alpha<0$ and the sufficiently large magnitude of $\alpha$, we can describe the color superconductivity with $N_c=3$. Estimate by numerical method in \cite{nam2021more}, we have the CSC occur at $N_c=2$ with $\alpha$ from $-4.2$ to $-2.0$ and at $N_c=3$ with $\alpha$ from $-9.0$ to $-6.5$   
\subsection{CSC phase in the confinement phase}
Now, we consider the case of confinement phase. To find the condensation of the diquark in the confinement phase, we need the condensate of the scalar field $\psi$ in the bulk which dual to the confinement phase in this model. The bulk correspond to the confinement phase is the GB-AdS soliton. And hence from\eqref{EOM soliton} we have the effective mass dual to the confinement phase
\begin{equation}
m^2_{eff soliton}=m^2-\frac{q^2\phi^2(r)}{r^2}
\end{equation}
To obtain the condensate we need the instability of the scalar field with the effective mass, it corresponds to the BF bound be broken:
\begin{equation}
\frac{q^2\phi^2(r)}{r^2}>\frac{9}{l^2_{eff}},
\end{equation}
with $\phi(r)$ given in \eqref{gauge soliton solution} we obtain
\begin{equation}
\frac{q\mu}{r_0}>\frac{3}{2}\sqrt{\frac{1-\sqrt{1-4\alpha}}{2\alpha}}
\end{equation}
The critical chemical potential of CSC phase in confinement phase must be smaller than $1.73$. In \cite{nam2021more}, we have the CSC phase in the confinement phase appear when $\alpha=-9.0$ or $\alpha=-8.0$ with $N_c=3$

\section{CSC phase equation of state \label{sec4}}
In the inner core of the massive neutron star, the phase transition will occur in which matter will transit from the baryon matter to the quark matter or the color superconductivity in the confinement phase. In this section, we will discuss the color superconductivity in the case of the deconfinement phase (finite temperature) and in the confinement phase (very low temperature in the cold compact star).  
\subsection{Equation of state in the deconfinement phase}
First of all, we will consider the color superconductivity in the deconfinement phase. In deconfinement phase, the color superconductivity phase transition will occur when the chemical potential (density) is large enough and hence the temperature is low. If we want to find the equation of state of the CSC phase by the EGB gravity holographic model, we need to compute the free energy of the color superconductivity phase in the deconfinement phase when  the diquark condensate appear. To calculate the free energy, we must calculate the on-shell action of GB-RN-AdS black hole. The Euclidean action is separate to the gravity part and the matter part as:
\begin{equation}
S^E=-\int d^{d+1}x\sqrt{-g}\mathcal{L}+S_{bnd}=S^E_{grav}+S^E_{ matter}.
\end{equation}  
From \cite{Olea2011} the gravity part of this Euclidean action is given by
\begin{equation}
\begin{split}
S^E_{grav}&=\left[ (r^2f)'(r^4-4\alpha r^4 f)\lvert^{\infty}_{r_+}-l^4_{eff}\left(1-\frac{4\alpha}{\l^2_{eff}}\right)r^4f^2(r^2f)'\lvert^{\infty} \right]\frac{4\pi}{5r_0}\frac{V_3}{T}\\
&=\hat{S}^E_{grav}\frac{4\pi}{5r_0}\frac{V_3}{T},
\end{split}
\end{equation}

with $l^2_{eff}=\frac{2\alpha}{1-\sqrt{1-4\alpha}}$.

We have
\begin{equation}
\hat{S}^E_{grav}=-r^5_+ +\frac{21\mu^2r^3_+}{8}.
\end{equation}
The matter part of this action $S^E_{matter}$ consists of three parts
\begin{equation}
S^E_{matter}=(\hat{S}^E_{\psi}+\hat{S}^E_{\phi}+S^E_{bnd,F})\frac{4\pi}{5r_0}\frac{V_3}{T}.
\end{equation}
For the first term, we have:
\begin{equation}
\begin{split}
\label{Spsi}
\hat{S}^E_{\psi}&=-\int dr\sqrt{-g}(-|D_{\mu}\psi|^2-m^2|\psi|^2)\\
&=\int dr\sqrt{-g}(g^{rr}\psi'^2+q^2A_0^2\psi^2g^{00}+m^2\psi^2)\\
&=\int dr\sqrt{-g}\left[-\frac{1}{\sqrt{-g}}\partial_r(\sqrt{-g}(g^{rr}\psi'))+q^2A^2_0\psi g^{00}+m^2\psi\right]\psi\\
&+[\sqrt{-g}g^{rr}\psi'\psi]^{\infty}_{r_+}.
\end{split}
\end{equation}
The integral part vanished by the equation of motion $\frac{1}{\sqrt{-g}}\partial_r(\sqrt{-g}(g^{rr}\psi'))=q^2A^2_0\psi g^{00}+m^2\psi$. By the boundary term, Eq.\eqref{Spsi} becomes:
\begin{equation}
\hat{S}^E_{\psi}=[\sqrt{-g}g^{rr}\psi'\psi]^{\infty}_{r_+}=[r^6f(r)\psi\psi']^{\infty}_{r_+}=0.
\end{equation}
Because $f(r_+)=0$ and $\psi(r)|_{r\rightarrow\infty} =\frac{C}{r^4}+...$

For the second term, we see:
\begin{equation}
\label{Sphi}
\begin{split}
\hat{S}^E_{\phi}&=-\int dr\sqrt{-g}\left(-\frac{1}{4}F^2\right)\\
&=-\int dr\sqrt{-g}\left(-\frac{1}{2}g^{00}g^{rr}\phi'^2\right)\\
&=-\frac{1}{2}\int dr\sqrt{-g}\left[\frac{1}{\sqrt{-g}}\partial_r(\sqrt{-g}g^{00}g^{rr}\phi')\right]\phi+\frac{1}{2}g^{00}g^{rr}\sqrt{-g}\phi\phi'\\
&=-\frac{1}{2}\int\sqrt{-g}2q^2g^{00}A_0^2\psi^2+\frac{1}{2}g^{00}g^{rr}\sqrt{-g}\phi\phi'\\
&=-\int\sqrt{-g}q^2g^{00}A_0^2\psi^2+\frac{1}{2}g^{00}g^{rr}\sqrt{-g}\phi\phi'.
\end{split}
\end{equation}
With the AdS black hole case, we have:
\begin{equation}
\hat{S}^E_{\phi}=\int^{\infty}_{r_+}dr\frac{q^2r^2\psi^2\phi^2}{f(r)}-\frac{1}{2}r^4\phi\phi'|^{\infty}_{r_+},
\end{equation}
and the boundary field action is
\begin{equation}
S_{bnd,F}=\frac{1}{2}\frac{4\pi}{5r_0}\frac{V_3}{T}\sqrt{|h|}n_aF^{ab}A_b|^{\infty},
\end{equation}
$h$ is the determinant of the induced metric at the boundary and $n_a=g_{aa}n^a$ ($n^a$ is a normal vector). We have
\begin{equation}
|h|=-g_{00}g_{11}g_{22}g_{33}g_{yy}=r^{10}f(r),
\end{equation}
and
\begin{equation}
n^a=\frac{1}{\sqrt{g_{rr}}}\left(\frac{\partial}{\partial r}\right)^a=\frac{\delta^a_r}{\sqrt{g_{rr}}}.
\end{equation}
We have:
\begin{equation}
n_r=g_{rr}n^r=g_{rr}\frac{1}{\sqrt{g_{rr}}}=\sqrt{g_{rr}}=\frac{1}{r\sqrt{f(r)}}.
\end{equation}
Hence, we obtain:
\begin{equation}
S_{bnd,F}=-\frac{1}{2}\frac{4\pi}{5r_0}\frac{V_3}{T}r^4\phi\phi'|^{\infty}=-\frac{4\pi}{5r_0}\frac{V_3}{T}\frac{3}{2}\mu\overline{d}.
\end{equation}

And we have the Euclidean action:
\begin{equation}\label{euclidean action black hole}
S^E=\frac{4\pi}{5r_0}\frac{V_3}{T}\left[-r_+^5+\frac{21\mu^2}{8}r^3_+ -3\mu\overline{d}+\int^{\infty}_{r_+}\frac{q^2r^2\phi^2\psi^2}{f(r)}dr\right].
\end{equation} 
Hence the free energy density
\begin{equation} \label{free energy}
\Omega=-r^5_++\frac{21\mu^2}{8}r_+^3-3\mu\overline{d}+\int^{\infty}_{r_+}\frac{q^2r^2\phi^2\psi^2}{f(r)}dr,
\end{equation}
and the pressure
\begin{equation} \label{the pressure}
p=-\Omega=r_+^5-\frac{21\mu^2}{8}r^3_+ +3\mu\overline{d}-\int^{\infty}_{r_+}\frac{q^2r^2\phi^2\psi^2}{f(r)}dr.
\end{equation}


The CSC phase in the boundary appears because of condensation of the diquark. This condensation breaks the $U(1)_B$ of the baryon symmetry. In this holographic model, it corresponds to the spontaneously broken $U(1)$ symmetry in the bulk. In this model, the $U(1)$ charge is kept fixed, the condensation of the scalar field in the bulk is triggered by the chemical potential, and this chemical potential corresponds to the baryon chemical potential asscociated with the density of matter in the inner core of a massive compact star (in case of hot compact star) \cite{nam2021more}. At the critical chemical potential $\mu=\mu_c$ the CSC phase transition occurs (the role of the critical chemical potential, $\mu_c$, analogy to the critical temperature $T_c$ in metallic superconductivity). However, in  the holographic CSC phase, we use the critical chemical potential because even $T=0$ if $\mu>\mu_c$ the CSC phase also appear and hence the CSC phase is characterized by $\mu_c$ instead of the critical temperature $T_c$ in metallic superconductivity. In the limit of $\mu=\mu_c$, the back reaction of the bulk scalar field is negligible. At $\mu=\mu_c$, $\psi=0$ and we have
\begin{equation}
 \phi(r)=\mu\left(1-\frac{r_+^3}{r^3}\right).
\end{equation} 

Above the critical chemical potential $\mu>\mu_c$, because the condensation of the diquark which corresponds to non-trivial bulk scalar fields $\psi$. The condensation of the pairs of quarks breaks the $SU(3)_C$ symmetry and this corresponds to the spontaneous breaking of the $U(1)$ symmetry in the bulk. Because the $U(1)$ symmetry is broken spontaneously in the bulk theory, we have $J_C=0$ and $C\neq 0$ from \eqref{boundary psi} we see that the asymptotic behavior of the bulk scalar field $\psi(r)$ becomes:
\begin{equation}\label{asymptotic psi}
    \psi(r)=\frac{C}{r^4}.
\end{equation}

In near $\mu_c$ (above but near) $\psi\neq 0$ but small enough to the back reaction of the bulk scalar field is negligible, and hence the field $\phi(r)$ is considered that it does not change. We obtain:
\begin{equation}
\begin{split}
\Omega&=-r_+^5+\frac{21\mu^2}{8}r_+^3-3\mu^2r_+^3+\int^{\infty}_{r_+}\frac{q^2r^2\phi^2\psi^2}{f(r)}dr\\
&=-r^5_+ -\frac{3\mu^2r^3_+}{8}+\int^{\infty}_{r_+}\frac{q^2r^2\phi^2\psi^2}{f(r)}dr,
\end{split}
\end{equation} 
and the pressure of the color superconducting gas in the inner core of the compact star is given by 
\begin{equation}\label{pressure via r}
p=-\Omega=r^5_+ +\frac{3\mu^2r^3_+}{8}-\int^{\infty}_{r_+}\frac{q^2r^2\phi^2\psi^2}{f(r)}dr.
\end{equation} 
Solving the equation of motion \eqref{eom} to estimate $\psi$ by the Sturn$-$Liouville method when $\mu\approx\mu_c$ \cite{nam2019} with $r_+=1$, by the boundary condition for the matter field \eqref{asymptotic psi} we have $\psi=Cz^4H(z)$ with $z=r_+/r$ and the trial function $H(z)$ is chosen by $H(z)=1-az^2$ \cite{nam2019}.

From vthe variable $z$ the equation \eqref{pressure via r} is rewritten by
\begin{equation}
p=1+\frac{3\mu^2}{8}-\frac{q^2\mu^2(1-z^3)^2C^2(1-az^2)^2z^4}{f(z)}dz
\end{equation}
where $f(z)=\frac{1}{2\alpha}\left(1-\sqrt{1-4\alpha(1-z^5)+\frac{3\alpha\mu^2}{2}z^5(1-z^3)}\right)$

And we have the equation of state $p=p(\mu)$
\begin{equation}
\label{EOS near critical}
\begin{split}
    p&=1+\frac{3\mu^2_c}{8}+\frac{3\mu_c(\mu-\mu_c)}{4}+\frac{3}{8}(\mu-\mu_c)^2-2\alpha q^2\int^1_0dzA(\mu_c,z)\\
    &-2\alpha 2q^2\int^1_0dzA'(\mu_c,z)(\mu-\mu_c)-\alpha 2q^2\int^1_0dzA''(\mu_c,z)(\mu-\mu_c)^2-... .
    \end{split}
\end{equation}
Here $A(\mu,z)=\frac{\mu^2(1-z^3)^2(1-az^2)^2z^4}{1-\sqrt{1-4\alpha(1-z^5)+\frac{3\alpha\mu^2}{2}z^5(1-z^3)}}$ and $A'(\mu,z)=\frac{\partial A(\mu,z)}{\partial\mu}$ and we can set $C=1$. In comparision with the baryon phase equation of state in \cite{Kazuo 2021} we find that the CSC phase is softer than the baryon phase of the compact star.
\subsection{Equation of state in the confinement phase}
The second case is the color superconductivity appear in the confinement phase. This case occur in the inner core of the cold compact star because the diquark can be formed in confinement phase. In \cite{nam2021more}, if we have the negative $\alpha$ and sufficient large magnitude we can describe the color superconductivity with in the confinement phase. Calculate analogously to the case of AdS black hole. We obtain the free energy density for the AdS soliton when the scalar hair (near critial chemical potential) appear:
\begin{equation}
\Omega_{Sol}=-r_0^5+\int^{\infty}_{r_0}\frac{q^2r^2\phi^2\psi^2}{f(r)}dr
\end{equation}
Hence the pressure is given by
\begin{equation}
p_{confinement}=-\Omega_{soliton}=r_0^5-\int^{\infty}_{r_0}dr\frac{q^2r^2\mu^2\psi^2}{f(r)}
\end{equation}
To solve the $\psi$ from the equation of motion \eqref{EOM soliton} with $r_0=1$, we introduce the variable $z=\frac{r_0}{r}$ and also use the Sturm-Liouville method \cite{nam2019}. From the boundary condition when $r \rightarrow\infty$ we have $\psi(z)=Cz^4H(z)$ and the trial function $H(z)=1-az^2$. The equation of state is given by
\begin{equation}
p=1-\int^1_0\frac{q^2\mu^2C^2(1-az^2)^2z^4}{f(z)}dz
\end{equation} 
where $f(z)=\frac{1}{2\alpha}(1-\sqrt{1-4\alpha(1-z^5)})$ and we also set $C=1$. Hence the equation of state of the color superconductivity in the confinement case is rewritten by
\begin{equation}
p=1-2\alpha q^2B\mu^2
\end{equation}
where $B=\int^1_0\frac{(1-z^2)z^4}{f(z)}dz$. It is also softer than the baryon phase equation of state in \cite{Kazuo 2021}




\section{Discussion \label{sec5}}
By the holographic model from Einstein-Gauss-Bonnet gravity, we computed the equation of state of the color superconductivity phase in the inner core of the massive compact star with both the cold core (corresponds to color superconductivity transition in the confinement phase) and hot core (correspond to the CSC phase in the deconfinement phase) of the compact star. Because the confinement-deconfinement phase transition occur when $\mu=1.73$, color superconductivity phase transition in the confinement phase appear when the chemical potential$\mu<1.73$ and in the deconfinement phase the color superconductivity phase transition appear when $\mu>1.73$. Near the critical point of the CSC  the equation of state \eqref{EOS near critical} shows that the CSC phase in the inner core is softer than the baryon matter in the crust in both confinement and deconfinement case. In the next step, we will solve the TOV equation \cite{TOV} to obtain the mass, radius and stability of this type of compact star. 

In the future, we will study this with the $p$-wave and the $d$-wave CSC phase (case of diquark is the vector field), the Meissner effect (consist of electromagnetic and color Meissner effect) of the CSC phase via holography,and we will study the holographic entanglement entropy of the CSC phase.

\section*{Acknowledgment}
We would like to thank Nam H. Cao for his helpful feedback and insightful discussions. Additionally, we are grateful to Dmitry Voskresensky for his valuable discussions on color superconductivity and neutron star physics.

\end{document}